\documentclass{iopart}  
\usepackage{epsfig,subfigure}
\usepackage{citesort}
\newcommand{\DIR}{figures}

\begin{document}
\jl{1}

\def\NEW#1{{\bf{#1}}\marginpar{$\longleftarrow$ {\bf NEW!}}}
\def\ADD#1{{\bf{#1}}\marginpar{$\longleftarrow$ {\bf ADD!}}}
\def\WOLFGANG#1{{\bf{#1}}\marginpar{$\longleftarrow$ {\bf WOLFGANG!}}}

\title{Optimization potential of a real highway network: an empirical study}

\author{Wolfgang Knospe$^1$, Ludger Santen$^2$, Andreas Schadschneider$^3$ 
and Michael Schreckenberg$^1$}

\address{$^1$ Theoretische Physik, Fakult\"at 4,
          Gerhard-Mercator-Universit\"at Duisburg,
                  Lotharstr. 1, 47048 Duisburg, Germany}

\address{$^2$ Fachrichtung Theoretische Physik, Universit\"at des
Saarlandes, Postfach 151150, 66041 Saarbr\"ucken, Germany}

\address{$^3$ Institut f\"ur Theoretische Physik,
                 Universit\"at zu K\"oln,
                 Z\"ulpicher Str. 77, 50937 K\"oln, Germany }

\date{\today}


\begin{abstract}
Empirical observations and theoretical studies indicate that the
overall travel-time of vehicles in a traffic network can be optimized
by means of ramp metering control systems. Here, we present an 
analysis of traffic data of the highway network of
North-Rhine-Westfalia in order to identify and
characterize the sections of the network which limit the performance,
i.e., the bottlenecks. It is
clarified whether the bottlenecks are of topological nature or if they are
constituted by on-ramps. This allows to judge possible optimization
mechanisms and reveals in which areas of the network they have to be
applied. 
\end{abstract}

\section{Introduction}

In the last few years, the increasing amount of vehicular traffic has led to a
more and more oversaturated freeway network. In contrast to the past,
 it is no longer possible to react on this additional amount of traffic
by construction of new highways.
Therefore, in order to use the existing network more efficiently,
several control 
strategies like signal control, variable message signs, route
guidance, motorway-to-motorway control and ramp
metering (Papageorgiou, 1995) have been proposed for the optimization
of the highway's performance and for the amelioration of the system's
traffic state. 

Empirical studies of parts of a highway network (Salem, 1995) have
revealed the benefits of ramp metering strategies for the total mean speed
and the total travel-time for both, the vehicles on the highway {\it and}
the queuing vehicles on the on-ramp.
These empirical findings are underscored by several theoretical
investigations of simplified (Kolomeisky, 1998; Appert, 2001) as 
well as well as of more realistic (Popkov, 2001; Huberman, 1999; 
Barlovic, submitted) traffic models. In 
particular, studies of the asymmetric simple exclusion process 
reveal the existence of a maximal current phase which obviously is the most
desirable state. It can be obtained for large input and
small output rates and allows to operate a highway in the 
optimal regime.  The possible capacity gain is even 
higher in more realistic models which reproduce the empirically 
observed meta-stable high-flow states. In finite systems these 
states are temporally quite stable and, therefore, it seems 
to be possible to optimize at least short highway sections in 
this manner.

Moreover, in (Huberman, 1999), a simple simulation setup consisting
of a two-lane highway segment 
with an on-ramp is presented. It has been shown that the
travel-time of all vehicles can be optimized at a finite injection
rate of the on-ramp in the sense that its variance is minimized. 

This improvement of the system's throughput can be explained by
considering fluctuations of the traffic flow. Obviously, free flow is
the most desirable traffic state since large flows can be
obtained. However, at increasing traffic demand, synchronized traffic
forms in the vicinity of on-ramps, which enables a large amount of
vehicles to drive with a  velocity considerably larger than in a 
jam but smaller than in free flow, and leads to a flow
comparable to free flow (Kerner, 2001). 
Synchronized states are not very stable,  i.e., already  
small perturbations can cause wide jams. The importance of 
synchronized traffic for an optimized usage of the network 
capacity is due to its role as a precursor of traffic jams: 
If synchronized traffic is observed at a bottleneck
control mechanisms will help to stabilize 
this phase and, thus, ensure large throughputs of the
highway at large densities. 
In particular, by means of ramp metering, it is possible to tune the
injection rate 
of the on-ramp in order to affect the vehicle-vehicle interactions. 
Increasing the interactions will improve the vehicle coupling and help to
synchronize them (Ariaratnam, 2001), however, strong interactions
lead to perturbations 
of the flow and, thus, to wide jams. The variation of the on-ramp flow
therefore helps to adjust the interactions and to stabilize the
synchronized state. As a result, the perturbations are minimized, and
the performance of the highway is optimized.

In this study we present an extensive analysis of a highway network
 which gives insight about the spreading of jams in the 
network. In contrast to
former analyses which are restricted to only small parts of a highway
system, here, the coverage of the network with inductive loops allows
to analyse the overall traffic state on a global scale. This enables
us to identify and characterize its bottlenecks. In particular, the
question is raised whether bottlenecks are of topological nature or if
they are 
constituted by on-ramps with a large inflow. 
Topological bottlenecks are static whereas dynamical bottlenecks (like
ramps) in principle can be influenced externally, e.g., by controlling
their strength.
Moreover, the properties
of the bottlenecks may give evidence if  
the network's state can be optimized by local control devices only 
or whether global traffic management strategies are 
necessary (Brockfeld, 2001).

\begin{figure} 
\begin{center}
\includegraphics[width=0.63\linewidth]{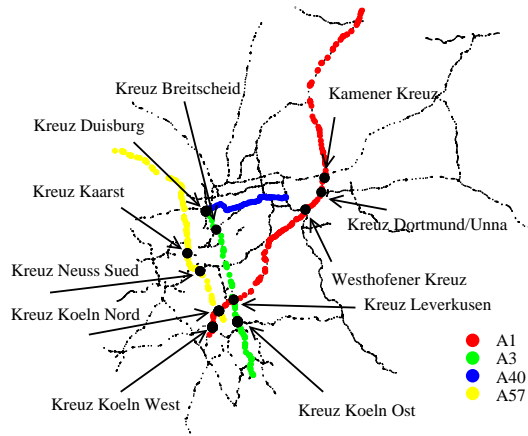}
\caption{Sketch of the highway network of North
Rhine-Westfalia and position of the main highways, the A1, the A3, the A57 and
the A40.}
\label{nrw}
\end{center}
\end{figure}

\begin{figure} 
\begin{center}
\includegraphics[width=0.63\linewidth]{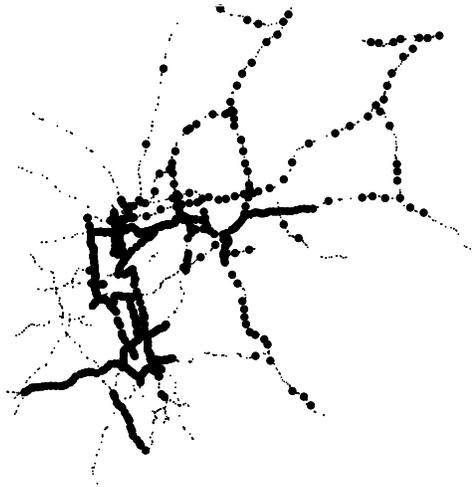}
\caption{Position of the inductive loops (black dots). As one can
see, the inductive loops concentrate on the inner region of the
network where most of the traffic volume can be found.}
\label{nrw_loops}
\end{center}
\end{figure}

\section{Network characteristics}

We examine the highway network of North Rhine-Westfalia (see
Fig.~\ref{nrw} for a sketch of the network) which has
a total length of
about $2~500~km$ and includes $67$ highway intersections and $830$ on- and
off-ramps. The data set is provided by about $3~500$ inductive loops
which send minute aggregated data of the flow, the occupancy and the
velocity online (that is every minute) via permanent lines from the
traffic control centers in Recklinghausen and in Leverkusen. For the
sake of simple data handling, the driving directions are classified
into south/west and north/east.

As one can see in Fig.~\ref{nrw_loops}, most of the detectors are
concentrated on the inner network. This is justified by the
distribution of the traffic volume, as we will show later.
The
recording of the data started on 10-10-2000. For the analysis of the
network, a period of $265$ days from 10-10-2000 to 07-01-2001 was considered.

\begin{figure}[h] 
\begin{center}
\includegraphics[width=0.63\linewidth]{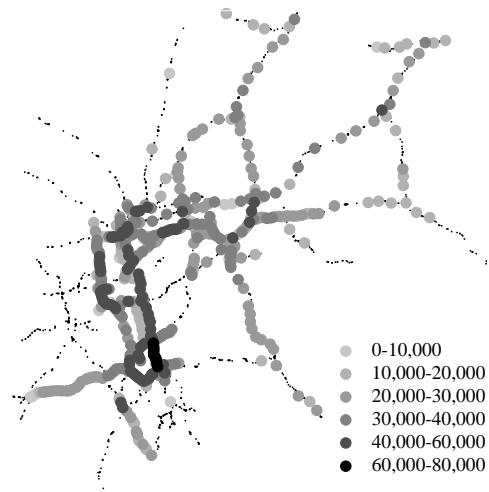}
\caption{Number of vehicles per measurement location per $24$
hours and driving direction.}
\label{kfz}
\end{center}
\end{figure}

The highway network of North Rhine-Westfalia has an average traffic
load of about $30~000$ veh/24h per measurement section for one
driving direction. Note that the traffic volume is calculated 
per measurement section and thus sums the flow of two or 
three lanes. However, there
are large differences between the highways (Fig.~\ref{kfz}). 
Obviously, only a few sections with very large traffic
volumes exist which concentrate on the main urban areas. In detail,
the K\"olner Ring, especially the section between the highway
intersections Kreuz Leverkusen and Kreuz K\"oln Ost, large parts of
the A3 and the 
A40, the A57 near Neuss and the A2 near the highway intersection Kreuz
Duisburg have an average load of $40~000-80~000$ veh/24h.
The more the highways leave these conurbations, the less is the traffic
volume which can be measured. 
In addition, nearly the same values of the traffic volumes can be
found for both, the driving directions south/west and north/east.

\begin{figure}[h] 
\begin{center}
\includegraphics[width=0.63\linewidth]{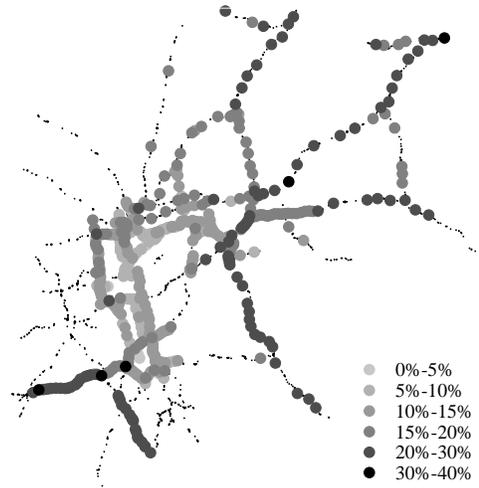}
\caption{Relative number of trucks per measurement section.} 
\label{lkwshare}
\end{center}
\end{figure}

About $15\%$ of the vehicles are trucks. 
Fig.~\ref{lkwshare} shows the distribution
of the relative number of trucks per measurement section. 
Obviously, the share of trucks on the traffic volume 
increases with the distance to the conurbations, while  
the absolute number of trucks does not change significantly on the main
highways. Thus, the long-distance traffic is mainly dominated by
trucks, while at short distances commuter traffic leads to large
traffic loads of the network.

\section{Bottleneck localization}

Since the traffic volumes are simple integrated quantities, it is possible
that a large traffic volume can indicate that a two-lane road is often
jammed, while the capacity of a three-lane road is still not reached.
Therefore, in order to allow an evaluation of the traffic load, the
probability to find a jam is calculated. This has the advantage that
traffic volumes which are larger than the highway capacity can
be related directly to a large jam probability, and thus bottlenecks of
the highway network can clearly be identified.
Due to the size of the analysed data set it is necessary 
to introduce a simple criterion to identify jams. Our criterion 
is the following: If the density 
of at least one minute is larger than $50\%$ (the density is 
given as occupancy which is the percentage of time a detector 
is covered by a vehicle), a jam is supposed. This criterion 
was motivated by former empirical studies (Knospe, 2002)
 that indicated that 
densities higher than $50\%$ are typically neither observed 
in free-flow nor in synchronized traffic. The threshold value as
well as the chosen time interval are, of course, somehow arbitrary. 
Our results are, however, rather insensitive upon an increase of 
the density threshold. By contrast, longer time intervals may 
lead to problems, because jams are often not compact but 
include regions of free flow traffic. This implies that the 
density threshold has to be lowered if longer time intervals 
are used, but lower density threshold lead to problems in 
discriminating between jams and synchronized traffic.

The usage of the density as an indicator for a jam has some advantages
compared to the mean velocity and the flow. First, the flow as well as
the velocity depend on speed limits which vary in the
network. Therefore, it is not 
possible to distinguish whether a small velocity is due to a large
density or a speed limit.
Second, a small flow can refer to both a jam or just free flow
with only a few cars. The jam probability is therefore calculated as
the relation of the number of days a jam was found to the total number of
days of the observation period ($265$ days). 

\begin{figure}[h] 
\begin{center}
\includegraphics[width=0.63\linewidth]{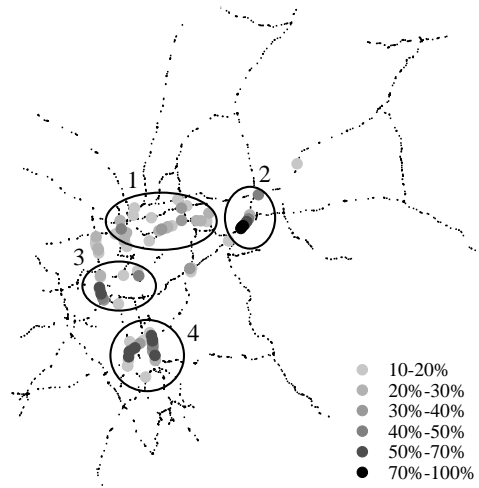}
\caption{Jam probability for the driving direction south and
west. The encircled regions are the Ruhrgebiet (region 1), the area of
Dortmund (region 2), the area of Krefeld and D\"usseldorf (region 3)
and the area of K\"oln (region 4).}
\label{pjam034}
\end{center}
\end{figure}

Fig.~\ref{pjam034} represents the jam probability of the
network for the driving direction south/west. Obviously, large jam
probabilities can be found in a region of the 
inner network, where the conurbations of the Ruhrgebiet (region 1), the areas of
Dortmund (region 2), D\"usseldorf and Krefeld (region 3) and K\"oln
(region 4) are located. 
As one can see in Fig.~\ref{nrw_loops}, most of the inductive loops are
concentrated in this regions, which  are indeed the most crowded
sectors of the network.
In contrast, in regions which are less equipped with 
counting devices, only few jams can be found.

Large differences of the jam probability
between the areas can be observed.
Only $10\%-30\%$ of the observation days were jammed in the area of
the Ruhrgebiet (region 1), that is every third day a jam occurred. In
contrast, in the areas of Dortmund (region 2), Krefeld and
D\"usseldorf (region 3) and K\"oln (region 4), more than $50\%$ of the
days showed 
jams. Especially in 
the area of Dortmund, between the Westhofener Kreuz and the Kreuz
Dortmund/Unna, nearly $5$ days a weak jams can be measured.

Next, we analyse {\it when} the jams typically occur. First, we 
expect significant differences between the weekend and 
working days. 

\begin{figure}[h] 
\begin{center}
\includegraphics[width=0.63\linewidth]{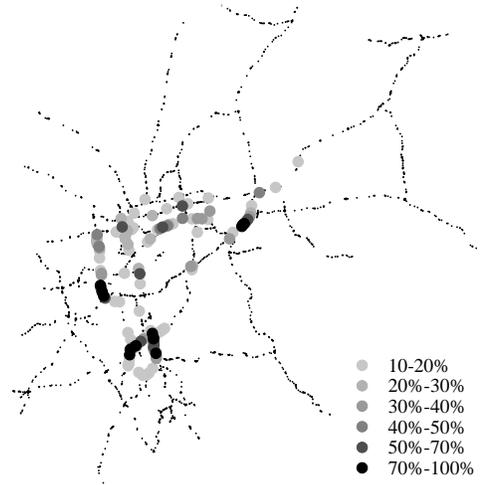}
\caption{Jam probability on wednesday for the driving
directions south and west.}
\label{pjam034_wednesday}
\end{center}
\end{figure}

Indeed, the exclusion of weekends leads to an increased jam
probability on weekdays compared to the 
probability averaged over the whole observation period.
 Second, we regarded the daily variations of the jam
probabilities. During the week from monday to friday
only minor differences in the jam probability exist, but 
one has to distinguish the period from monday to 
thursday and the friday as we will show below.
The jam probabilities we obtained are qualitatively 
the same as in case of an average over the whole
data set  (see Fig.~\ref{pjam034_wednesday}), 
the same four regions show a large jam probability
 that is the Ruhrgebiet and the areas
of Dortmund, Krefeld and D\"usseldorf and K\"oln.
The actual value of the jam probability, however, is increased and the 
difference  to the remainder of the network is now more pronounced.
During the weekend the jam probability is considerably 
reduced. Nevertheless, there are a few locations in the network where high jam 
probabilities can be observed on sundays, i.e., between the Westhofener
Kreuz and the Kreuz Dortmund/Unna. 
Remarkably the same classification scheme was found to be 
appropriate for city data as discussed 
in (Chrobok, 2000; Chrobok, 2001).

\begin{figure}[p] 
\begin{center}
\includegraphics[width=0.63\linewidth]{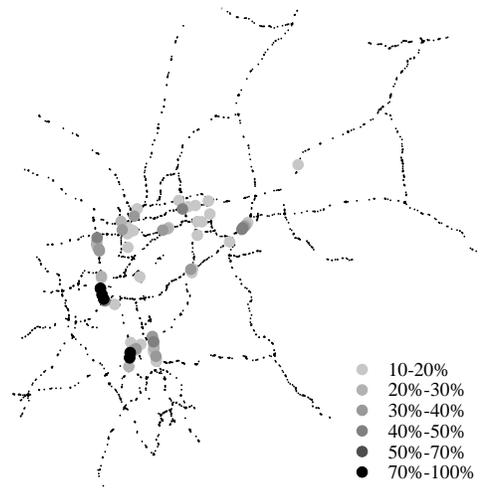}
\caption{Jam probability for the driving direction south and west
averaged over the whole observation period for the time interval
$6$-$11$ a.m.} 
\label{pjam034_611}
\end{center}
\end{figure}

 Next, we analyse the intraday variations of the data set.
We further divided the time-series into five time intervals, in order to
capture the main rush-hours. The first interval
covers the time from $0$ to $6$ a.m., the second interval from $6$ to
$11$ a.m., the third interval from $11$ a.m. to $1$ p.m., the fourth
interval from $1$ to $8$ p.m. and the fifth interval from $8$ to $12$
p.m. Indeed, the division of a day is somehow arbitrary, but 
a finer discretization of a day will not change the results
significantly. Moreover, analyses of city data revealed the existence
of basically two rush-hour peaks in the time-series of the traffic
volume at about $8$ a.m.~and $4$ p.m.~(Chrobok, 2000; Chrobok, 2001).

\begin{figure}[p]
\begin{center}
\includegraphics[width=0.63\linewidth]{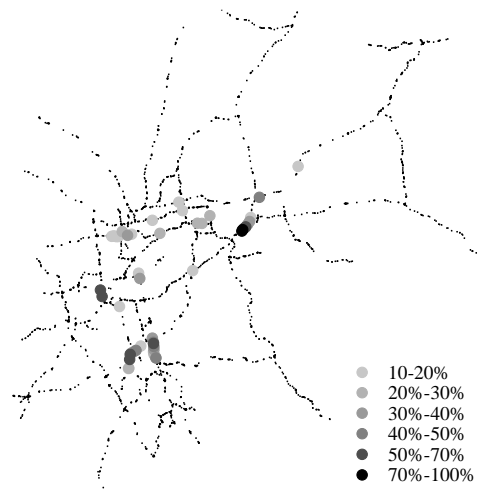}
\caption{Jam probability for the driving direction south and west
averaged over the whole observation period for the time interval $1$-$8$ p.m.}
\label{pjam034_1320}
\end{center}
\end{figure}

Figures~\ref{pjam034_611} and~\ref{pjam034_1320} 
show the jam probability of the whole data set in
the driving direction south/west for
two time intervals. During the first time interval, that is $0$ to
$6$ a.m., no jams could be measured. The main traffic volume occurs in
the second and in the fourth interval, which include the rush-hours. 
From $11$ a.m.~to $1$ p.m.~only a few jams are visible, while
from $8$ p.m.~to $12$ p.m.~just one region shows jams, namely the
region between the Westhofener Kreuz and the Kreuz Dortmund/Unna. 
The jams therefore occur mainly in the morning and afternoon
rush-hours.  There are, however, parts of the network where
the load is so large, that the probability to find a jam is high
throughout the whole day.

The same picture can be drawn from the analysis of the data set
classified in single days
which are divided into the five time intervals.
Due to the exclusion of the weekends, the values of the jam
probability increase significantly. 
The main disturbances of the network can be measured in the
rush-hours, while the regions $2$ and $4$ show continuously
a large traffic load. 
Again, there are only small differences between the weekdays. 
On fridays on the one hand, the load of the network in
the time between $6$ and $11$ a.m.~is smaller, but, on the other hand,
is 
increased significantly between $1$ and $8$ p.m.~compared to monday to
thursday. 
Moreover, on sundays, the large jam probability in the area of Dortmund
can be traced back to congestions between $1$ to $8$ p.m.

\begin{figure}[p] 
\begin{center}
\includegraphics[width=0.63\linewidth]{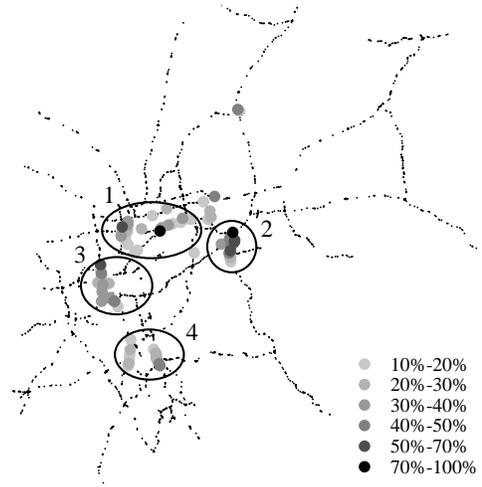}
\caption{Jam probability for the driving directions north and east
averaged over the whole data set. The encircled regions are the
Ruhrgebiet (region 1), the area of 
Dortmund (region 2), the area of Krefeld and D\"usseldorf (region 3)
and the area of K\"oln (region 4).}
\label{pjam002}
\end{center}
\end{figure}

Analogously to the analysis of the data of the driving direction
south/west, the jam probability is calculated for the
direction north/east (Fig.~\ref{pjam002}).
Like for the driving direction south/west, the
same regions with large jam probabilities occur which can be identified as
the conurbations of the Ruhrgebiet (region 1), the areas of Dortmund
(region 2), Krefeld and 
D\"usseldorf (region 3) and K\"oln (region 4).   
In contrast, the region of the Ruhrgebiet (region 1) and the area Krefeld and
D\"usseldorf (region 3) have a larger jam probability in the direction
north/east, while in 
the direction south/west the area of K\"oln (region 4) shows a
larger traffic 
volume. In the area of Dortmund (region 2), a large jam probability
can be found in the
south of the Westhofener Kreuz. 
Because north of the Westhofener Kreuz in the driving
direction south a large jam probability was measured, too, one 
can clearly identify the Westhofener Kreuz as a dominant bottleneck.

Again, the distinction between the days shows no difference to the
analysis of the whole observation period with the exception of a
larger jam probability in the single regions.
In addition, the main traffic volume can be measured in the morning
and afternoon rush-hours.

\section{Spatial extension of jams}

Calculating the jam probability in the four regions, one can observe
that consecutive measurement sections are often jammed even at the
same time, which may be a consequence of jams that have a large spatial
extension. However, as one can see further, a sequence of jammed
detectors is always restricted by highway intersections, not by on-
and offramps. But more importantly a jam which branches
from one highway to 
another highway via an intersection has been observed only in one case
(see below).

In order to determine the spatial extension of a jam,
it is necessary to consider its temporal occurrence:
 If for a given day a jam was detected at a detector, the 
detector is considered to be active (1) if not as being passive (0).
In this way we generated 
a binary time-series 
which allows the calculation of the spatial correlation between
different measurement locations. The explicit calculation
of the 
correlation of the density time-series, in contrast, has the
disadvantage that the 
duration of jams decreases in upstream direction,
which leads to vanishing correlations. 

\begin{figure}[p] 
\begin{center}
\includegraphics[width=0.63\linewidth]{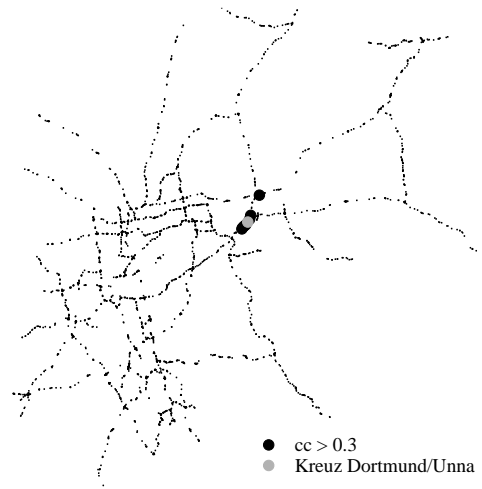}
\caption{Spatial correlation of the activity in the driving directions
south and west near the highway intersection Kreuz Dortmund/Unna. 
Correlation values larger than $0.3$ are shown.}
\label{cc034_4114074}
\end{center}
\end{figure}

\begin{figure}[p]
\begin{center}
\includegraphics[width=0.63\linewidth]{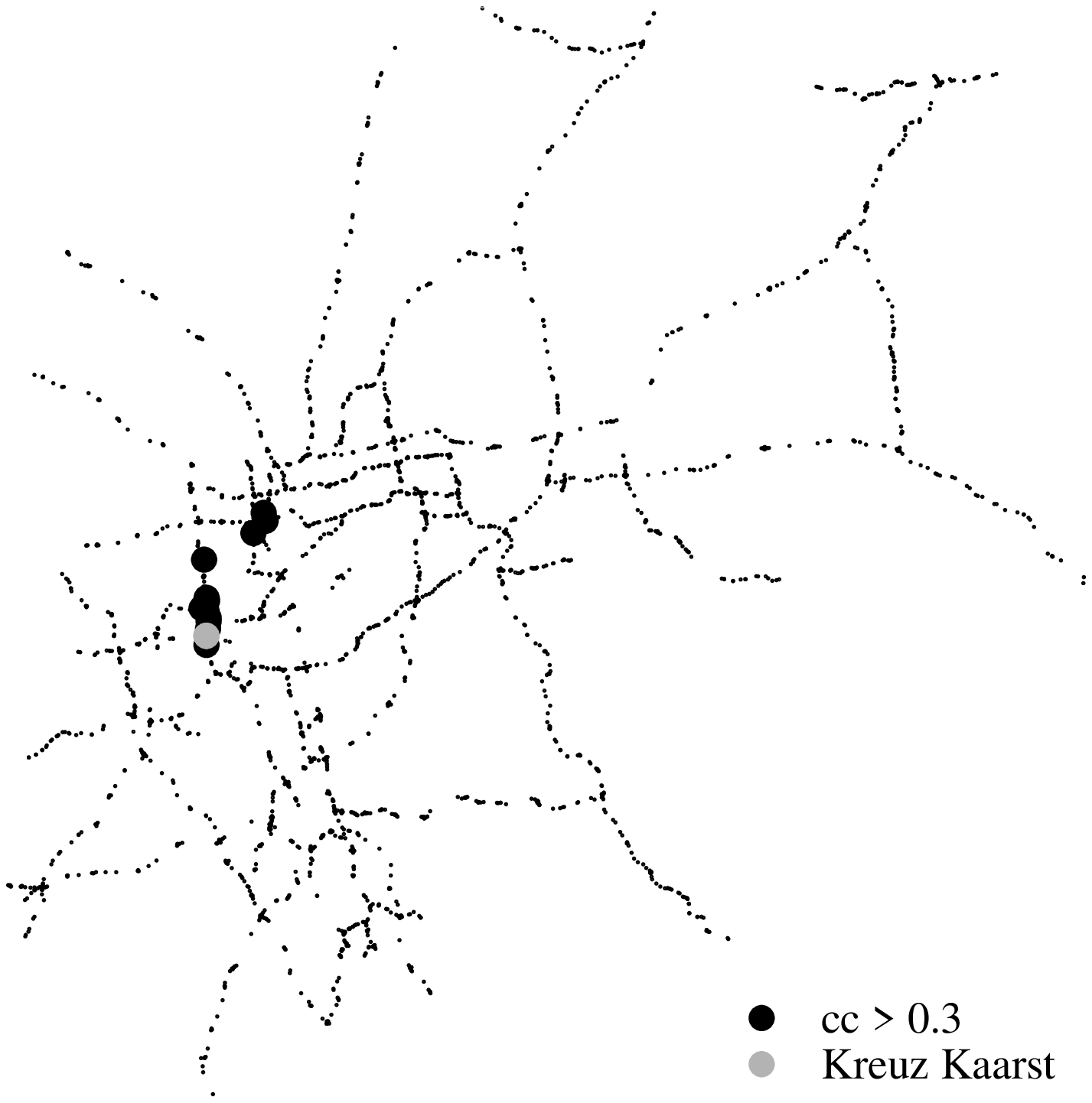}
\caption{Spatial correlation of the activity in the driving directions
north and east near the highway intersection Kreuz Kaarst. 
Correlation values larger than $0.3$ are shown.}
\label{cc002_6257028}
\end{center}
\end{figure}

Although the activity gives only a rough estimate of the jamming
state at an inductive loop, the typical spatial extension of jams
 can clearly be seen.
The large jam probability in region 2 in the driving direction
south/west can be related to a 
jamming of large parts of the A1 between the Westhofener Kreuz and the
Kreuz Dortmund/Unna (Fig.~\ref{cc034_4114074}) since the correlation
shows large values.
The same picture can be drawn in the area of K\"oln (region 4) where
a jam may 
cover a distance from the Kreuz K\"oln West 
and from the Kreuz K\"oln Ost to the Kreuz Leverkusen.
The largest
extension of a jam can be found on the A40 (region 1) where strong correlations
between the Kreuz Duisburg and the Kreuz Bochum
are measured. 
Obviously, there are also correlations with highways that are very
far away from the reference detector, but which are not connected by a
sequence of succeeding inductive loops with a large jam probability. 
The correlations are simply a consequence of a large 
traffic load of the network which leads to many jams on various
highway sections.
Since the number of jams
counted at the 
reference detector is large, correlations of a free flow signal can be
excluded. 

\begin{figure}[p] 
\begin{center}
\includegraphics[width=0.63\linewidth]{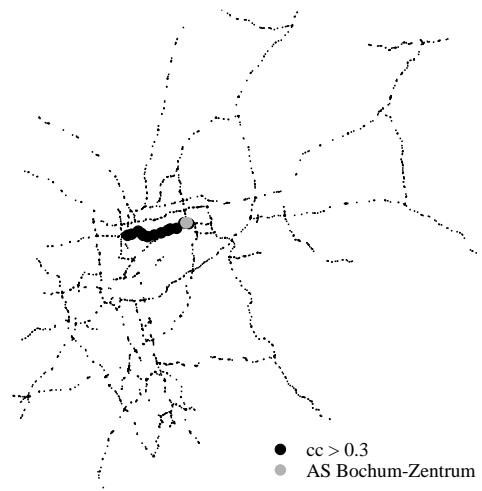}
\caption{Spatial correlation of the activity in the driving directions
north and east near AS Bochum-Zentrum. 
Correlation values larger than $0.3$ are shown.}
\label{cc002_5902002}
\end{center}
\end{figure}

\begin{figure}[p] 
\begin{center}
\includegraphics[width=0.63\linewidth]{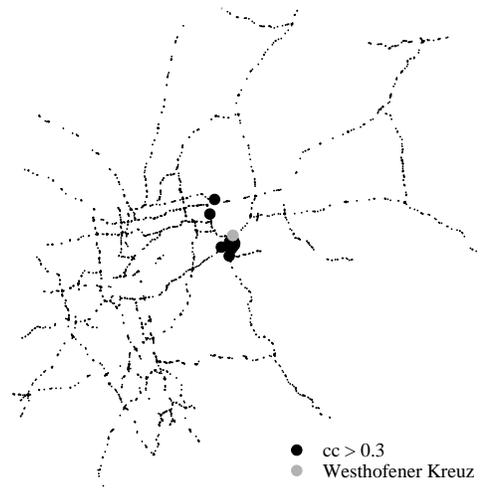}
\caption{Spatial correlation of the activity in the driving directions
north and east near the highway intersection Westhofener Kreuz. 
Correlation values larger than $0.3$ are shown.}
\label{cc002_4115017}
\end{center}
\end{figure}

The correlation analysis of the driving direction north/east reveals
strong values on the A57 near the Kreuz Kaarst
(Fig.~\ref{cc002_6257028}) and again on the A40 between the Kreuz Bochum
and the Kreuz Duisburg (Fig.\ref{cc002_5902002}). 
A jam measured at the Westhofener Kreuz (Fig.\ref{cc002_4115017}) can
branch due to the highways intersection on both, the highway A45 and
the highway A1.

The analysis of the activity allows for the daily occurrence of jams,
but temporal differences due to the morning or afternoon rush-hours
are omitted. Therefore, the activity is classified into the five
periods used for the jam probability calculation. This additional
consideration of the temporal occurrence of the activity allows a
better determination of the spatial correlation of the measurement
locations.

\begin{figure} 
\begin{center}
\includegraphics[width=0.63\linewidth]{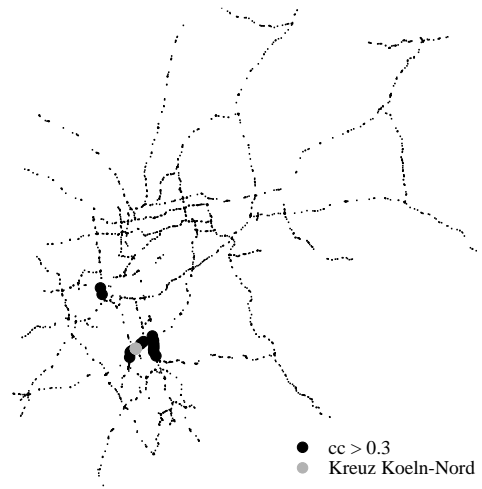}
\caption{Spatial correlation of the classified activity in the driving
directions 
south and west near the highway intersection Kreuz K\"oln-Nord. 
Correlation values larger than $0.3$ are shown.}
\label{cc034p_4133047}
\end{center}
\end{figure}

The picture of large
congested areas on the highways can be confirmed by the classified
activity correlation analysis. Especially the A40 between the Kreuz
Bochum and the Kreuz Duisburg in the
directions east and west, the A57
in the vicinity of the Kreuz 
Kaarst in the direction north and the area
of K\"oln in the direction south/west
(Fig.~\ref{cc034p_4133047}) show strong
correlations of 
consecutive measurement locations. 
Since the finer discretization of the activity time-series does
not change the results significantly, the classification of
the activity into five intervals is sufficient
for the proper recording of the temporal occurrence of jams.
Thus, although the classified activity is only a rough estimate for
the density time-series of a detector, the correlation analysis
nevertheless allows the determination of the spatial extension of jams.

\begin{figure} 
\begin{center}
\includegraphics[width=0.63\linewidth]{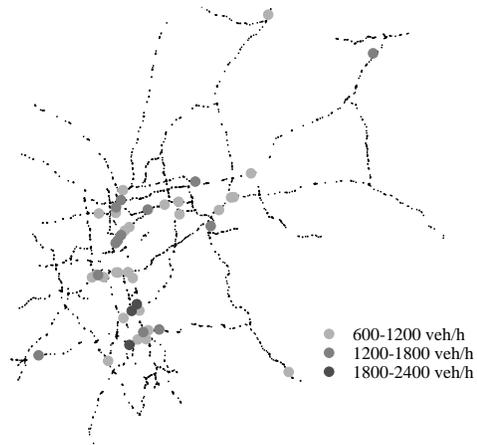}
\caption{Source rate in the driving directions south and west.}
\label{sources}
\end{center}
\end{figure}

\begin{figure} 
\begin{center}
\includegraphics[width=0.63\linewidth]{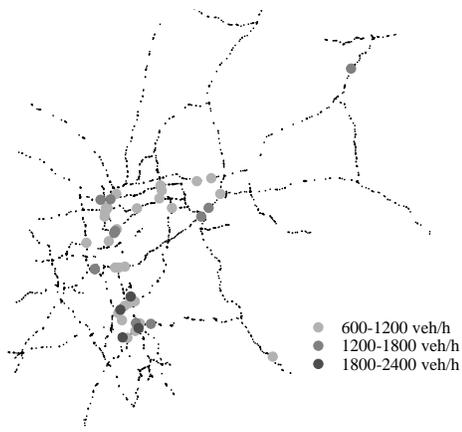}
\caption{Sink rate in the driving directions south and west.}
\label{sinks}
\end{center}
\end{figure}

The correlation analysis supports the results drawn from the jam
probability calculation. Although jams can have a large spatial
extension, branching of a jam via a highway intersection is rarely
observed.  As a consequence one can conclude that the bottlenecks are,
apart from the Westhofener Kreuz, a result of perturbations due to 
on- and offramps rather then local capacity restrictions.
As one can see in Fig.~\ref{sources} and Fig.~\ref{sinks}, most of the
sources of the network with a large 
injection rate are concentrated in the four regions. In addition, most
of the sinks can be found there~\footnote{The source and sink rate is
calculated by summing up the flow of all lanes at a measurement
section. The difference of the total flow between two succeeding
sections gives the loss or gain of the number of
vehicles.}. Therefore, traffic in these four 
regions is determined by strong fluctuations of the traffic demand
and, thus, the flow, resulting in jams with a large spatial
extension. However, these jams are restricted to single highways 
and do not affect traffic in other parts of the network.

\begin{figure} 
\begin{center}
\includegraphics[width=0.63\linewidth]{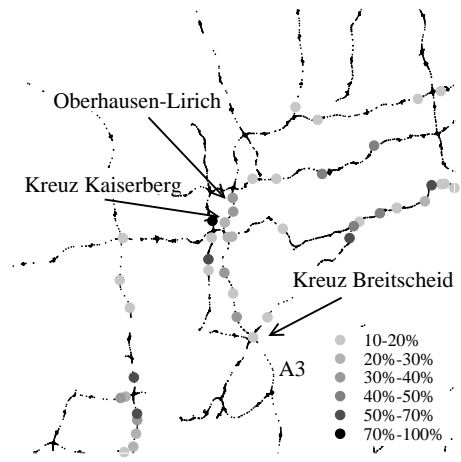}
\caption{Jam probability in the driving direction north and east in
October 2000 during the road constructions north of the junction
Oberhausen-Lirich.}
\label{lirich_before}
\end{center}
\end{figure}

\begin{figure} 
\begin{center}
\includegraphics[width=0.63\linewidth]{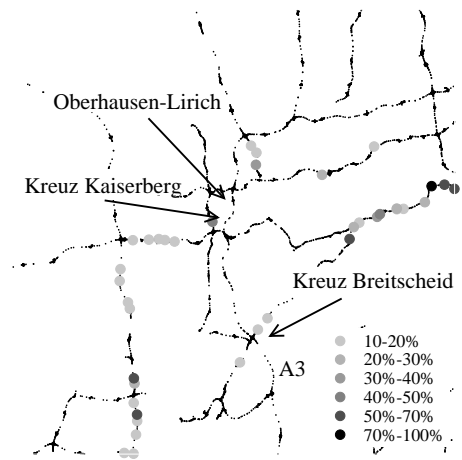}
\caption{Jam probability in the driving direction north and east in
October 2001 after finishing the road construction at
Oberhausen-Lirich.}
\label{lirich_after}
\end{center}
\end{figure}

Nevertheless, an example of a topological bottleneck is given in
Fig.~\ref{lirich_before}. The bottleneck is located north of the junction
Oberhausen-Lirich on the A3 and is generated by a reduction from three
to two lanes in each driving direction. In September 2001 this
bottleneck was eliminated by the extension to three lanes.
Due to the road works, jams regularly emerged in the south of the
bottleneck. These jams could have an extension up to the Kreuz
Breitscheid, passing the Kreuz Kaiserberg undisturbed. 
For comparison, Fig.~\ref{lirich_after} shows the same part of the
network after the road constructions have been finished. Obviously, the
probability to find a 
jam on the A3 south of the junction Oberhausen-Lirich is reduced
drastically. A systematical impact on other parts of the highway
network, however, cannot be observed.

\section{Conclusion}

The analysis of the highway network of North Rhine-Westfalia leads to
the following results: The main traffic load concentrates on the inner
network where the conurbations of the Ruhrgebiet (region 1), the areas of
Dortmund (region 2), D\"usseldorf and Krefeld (region 3) and K\"oln
(region 4) are located. In these regions, traffic is predominantly determined
by commuter or local traffic. In contrast, the outer regions of the
network are dominated by long-distance traffic, indicated by a
large truck share on the total traffic volume.
Jams can be measured very often in the central part of the network
i.e., jam probabilities of more than $50\%$ have been measured.
 This picture can be confirmed by a
more detailed analysis of the traffic data which takes the daily
occurrence of jams into account. Like in~(Chrobok, 2000; Chrobok, 2000),
the days can be classified into monday to thursday, friday, saturday
and sunday. However, the time a jam emerges is considered by the
subdivision of one day into five intervals. As a result, most of the
jams can be traced back on the morning and afternoon rush-hours, but
for the areas of K\"oln (region 4) and Dortmund (region 2), the
probability to find a jam is 
large during the whole day. In addition, on fridays, the main traffic
load is shifted from the morning to the afternoon rush-hour.

The correlation analysis of the activity takes the temporal occurrence
of a jam at consecutive measurement locations into account. Large
correlations between succeeding detectors in the four regions can be
measured. Nevertheless, 
the jams are mainly restricted to single highways and do not branch on
other highways via intersections. The only bottleneck which can be
identified is the highway intersection Westhofener Kreuz.
A more detailed calculation of the
correlation by means of a classified activity confirms these results.

The bottlenecks of the network are therefore predominantly on- and
offramps rather than topological peculiarities of the highway system. It is
the large in- or outflow at ramps that perturbes the stream of vehicles
on the main highway. As proposed in (Salem, 1995; Huberman, 1999), it
has to be expected that ramp metering systems are able to counteract
these destabilization of the flow and reduce the formation of jams.
In this way a restricted flow on the ramps may lead to a  
significant increase of the capacity of the main highways.
Of special interest for a capacity optimization is the 
existence of some ``hot spots'' in the network: The highest 
jam probabilities are spatially and temporally well localized. 
 These are the parts of the network, where the flow 
has to be optimized. Presently these sections self-organize in 
a congested state, leading fluxes that are far below their capacity. 
The situation can be improved a lot by controling the number 
of entering cars and by optimizing the traffic stream at the 
on- and off-ramps at this section.
The small number of bottlenecks that are present in the 
network shows that it is possible 
to improve the capacity of the network with a reasonable
technical effort. Of particular interest are sections where the 
main input is from other, less crowded, highways. In these 
cases a restricted input does not lead to a collapse of the
urban traffic. 

There is of course another way to avoid congestion in conurbations.
Our analysis shows that the jams are caused mainly by commuters.
This means that a better usage of public transport in the urban 
areas would indeed reduce significantly the congestion at the 
conurbations.

Summarizing our analysis has shown that the bottlenecks of a highway 
network can be found by means of a simple statistical analysis. 
The distribution of the bottlenecks in the network indicates that
optimization strategies can be successfully applied in order to increase 
the capacity of the network.

\vspace{0.2cm}

{\bf Acknowledgments}: 
The authors are grateful to the Landesbetrieb Stra\ss enbau NRW 
for the data support and to the 
Ministry of Economics and Midsize Businesses, Technology and Transport
as well as to the Federal Ministry of Education and Research of
Germany for the financial support (the latter within the BMBF project
``DAISY'').

\section*{References}
\begin{description}

\item[Appert, C.,] and L.~Santen. (2001).
"Boundary Induced Phase Transitions in Driven Lattice Gases with
 Metastable States." 
{\em Physical~Review~Letters}  86, 2498--2501. 

\item[Ariaratnam J.~T.,] and  S.~H.~Strogatz (2001).
"Phase Diagram for the Winfree Model of Coupled Nonlinear Oscillators."
{\em Physical~Review~Letters} 86, 4278--4281.

\item[Barlovic, R.,] T.~Huisinga, A.~Schadschneider,
and M.~Schreckenberg. (submitted).
"Open boundaries in a cellular automaton model for traffic
flow with metastable states."

\item[Brockfeld E.,] R.~Barlovic, A.~Schadschneider,
and M.~Schreckenberg. (2001).
"Optimizing Traffic Lights in a Cellular Automaton Model for City Traffic."
{\em Physical Review E} 64,  article number: 056132.

\item[Chrobok~R.,] O.~Kaumann, J.~Wahle,
and  M.~Schreckenberg. (2000). 
"Three Categories of Traffic Data: Historical, Current, and Predictive." 
In  E. Schnieder, and  U.~Becker (eds.),
{\em 9th IFAC Symposium Control in Transportation Systems.}
 Braunschweig: IFAC, pp.~250--255.

\item[Chrobok~R.,] J.~Wahle, and
  M.~Schreckenberg. (2001).
 "Traffic Forecast Using Simulations of Large Scale Networks."
 In B.~Stone, P.~Conroy, and A.~Broggi (eds.),
 {\em 4th International IEEE Conference on Intelligent 
 Transportation Systems.},
 Oakland: IEEE, pp.~434--439. 

\item[Huberman  B.A.,] and D.~Helbing. (1999).
"Economics-based optimization of unstable flows."
{\em Europhysics Letters } 47, 196--202.

\item[Kerner B.S.] (2001). 
"Complexity of Synchronized Flow and Related Problems for Basic
 Assumptions of Traffic Flow Theories." 
{\em Networks and Spatial Economics} 1, 35--76.

\item[Knospe~W.,] L.~Santen, A.~Schadschneider,
and M.~Schreckenberg. (2002).
"Single-vehicle data of highway traffic: microscopic description
 of traffic phases." 
{\em Phys.~Rev.~E}, in press. 

\item[Kolomeisky A.~B.,] G.~M.~Sch\"{u}tz,
E.~B.~Kolomeisky, and J.~P.~Straley. (1998).
"Phase diagram of one-dimensional driven lattice gases with open boundaries."
{\em  Journal of Physics ~A}  31, 6911--6919.

\item[Papageorgiou M.] (1995). 
"An Integrated  Control Approach for Traffic Corridors."
 {\em Transportation  Research C} 3, 19--30.

\item[Popkov V.,] L.~Santen, A.~Schadschneider,
G.~M.~Sch{\"u}tz.  (2001).
"Empirical evidence for a boundary-induced nonequilibrium phase
 transition."
{\em Journal of Physics A} 34, L45--L52.

\item[Salem H.~H.,] and M.~Papageorgiou. (1995).
   "Ramp Metring Impact on Urban Corridor Traffic: Field Results." 
   {\em Transportation Research A} 29, 303--319.

\end{description}

\end{document}